# Nearly Isotropic superconductivity in (Ba,K)Fe$_2$As$_2$

H. Q. Yuan[1,2], J. Singleton[2], F. F. Balakirev[2], S. A. Baily[2],

G. F. Chen[3], J. L. Luo[3], N. L. Wang[3]

[1] *Department of Physics, Zhejiang University, Hangzhou, Zhejiang 310027, China*

[2] *NHMFL, Los Alamos National Laboratory, MS E536, Los Alamos, NM 87545, USA.*

[3] *Beijing National Laboratory for Condensed Matter Physics, Institute of Physics, Chinese Academy of Science, Beijing 10080, China.*

**Superconductivity was recently observed in the iron-arsenic-based compounds with a superconducting transition temperature ($T_c$) as high as 56K[1,2,3,4,5,6,7], naturally raising comparisons with the high $T_c$ copper oxides. The copper oxides have layered crystal structures with quasi-two-dimensional electronic properties, which led to speculations that reduced dimensionality (that is, extreme anisotropy) is a necessary prerequisite for superconductivity at temperatures above 40 K[8,9]. Early work on the iron-arsenic compounds seemed to support this view[7,10]. Here we report measurements of the electrical resistivity in single crystals of (Ba,K)Fe$_2$As$_2$ in a magnetic field up to 60 T.  We find that the superconducting properties are in fact quite isotropic, being rather independent of the direction of the applied magnetic fields at low temperature. Such behaviour is strikingly different from all previously-known layered superconductors[9,11] , and indicates that reduced dimensionality in these compounds is not a prerequisite for "high-temperature" superconductivity.  We suggest that this situation arises because of the underlying electronic structure of the iron-arsenide compounds, which appears to be much more three dimensional than that of the copper oxides.**

**Extrapolations of low-field single-crystal data incorrectly suggest a high anisotropy and a greatly exaggerated zero-temperature upper critical field.**

(Ba,K)Fe$_2$As$_2$ was the first superconductor found in the (A,K)Fe$_2$As$_2$ (A=Ba, Sr) (or 122) series[6]; its superconducting $T_c$ can be as high as 38 K. Its parent compound BaFe$_2$As$_2$ shows a first-order structural phase transition from tetragonal to orthorhombic with the simultaneous onset of long-range antiferromagnetic order around 140 K[7,12]. Thermodynamic and transport measurements have been used to extract the $H_{c2}(T)$ dependences (where $H_{c2}$ is the upper critical field, and $T$ is temperature ) at low magnetic fields for fields applied both within the *ab* planes and parallel to the *c* axis. The estimated upper critical field $H_{c2}(0)$ based on these low-field $H_{c2}(T)$ experiments is anisotropic and exceeds 100 T at low temperatures[7]. However, experience with other quasi-two-dimensional superconductors such as organics and copper oxides shows that the low-field temperature dependence of $H_{c2}(T)$ is often a very poor guide to both its detailed behaviour at lower temperatures and its eventual $T = 0$ limit[9,11] Measurements in high magnetic fields are therefore very important, as they alone can give quantitative values of $H_{c2}(T\rightarrow0)$, thereby probing the mechanisms that limit the robustness of the superconducting state.

Fig. 1a shows the zero-field resistivity as a function of temperature for a single crystal of (Ba,K)Fe$_2$As$_2$ (This sample, denoted A, has the formula Ba$_{1-x}$K$_x$Fe$_2$As$_2$, where $x \approx 0.4$). A very sharp superconducting transition can be clearly observed with a midpoint at $T_c \approx 28.2$ K, indicating good sample quality. Within the normal state (either temperature- or field-induced), the Hall resistance follows a linear field dependence up to 60 T; this is in contrast to the underdoped ReFeAs(O,F) (1111-type) compounds in which the Hall resistance deviates from linearity at high fields and low temperatures [H. Q. Yuan, unpublished]. The Hall coefficient $R_H$ of (Ba,K)Fe$_2$As$_2$ is plotted as a function of temperature in the inset of Fig. 1a. Note that $R_H$ decreases monotonically with increasing temperature above $T_c$, consistent with an increase in the number of dominant hole carriers with increasing temperature.

The rather isotropic behaviour of $H_{c2}(T)$ at low temperatures that is the main focus of this paper is already evident in Figs. 1b and 1c; at the same temperature, superconductivity is suppressed by similar values of the field applied parallel or perpendicular to the $c$-axis. However, the normal state just above the critical field displays a distinct field orientation dependence: as the temperature decreases, the high-field resistance exhibits a slight "hump" for $H//c$, but decreases monotonically for in-plane magnetic fields. This is shown in greater detail in Fig. 2. (This figure shows data for samples A and B, both cut from the same batch with $x \approx 0.4$.) Similar phenomena occur in the underdoped 1111-type polycrystalline samples[13], high $T_c$ copper oxides[9] and organic metals[14]; in the latter case the "hump" is attributed to magnetoresistance caused by semiclassical electron trajectories across the Fermi surface[15].

Fig. 3 plots the resistive midpoint upper critical fields $H_{c2}$ versus temperature for two different samples of $(Ba,K)Fe_2As_2$ (#A and #B) which have almost exactly the same $T_c$ ($\approx 28K$) and resistive upper critical field behaviour. The most remarkable aspect of Fig. 3 is the fact that the resistive upper critical field of $(Ba,K)Fe_2As_2$ extrapolates to a similar zero-temperature value ($\approx 70$ T), irrespective of whether the field is applied parallel or perpendicular to the c axis. This is in great contrast to the behaviour of other quasi-two-dimensional superconductors such as the crystalline organic metals[11,14] or the copper oxides[9,16,17] where the in-plane critical fields are many times larger than those for fields applied perpendicular to the quasi-two-dimensional planes.

The magnetic-field-orientation dependence of the resistive upper critical field depends on two inter-related properties of the superconductor in question: (i) the underlying Fermi-surface topology and (ii) the nature of the vortices that form in the mixed state[9,14,16,17]. The Fermi-surface topology uniquely determines the velocities of the electrons involved in forming Cooper pairs and thus the ability of the material to support the circulatory currents involved in forming vortices[9,14,16,17]. Topologically, circulatory currents are possible if the intersections of planes perpendicular to the applied field and the Fermi surface produce closed loops[11]. If such currents occur, the additional "orbital energy" accumulated by the vortex system eventually destroys the superconducting state as the applied field increases.

In the case of the layered copper oxides and organic superconductors, the Fermi surfaces are very two-dimensional, with a cross-sectional area that varies little in the interlayer direction[9,14,18]. Hence, fields applied exactly within the conducting planes cannot induce significant circulating currents, as the Fermi surface cross-sections perpendicular to this are almost all open[15]. This prevents orbital mechanisms from limiting the upper critical field[14,16,17]. In such cases, the superconductivity is suppressed at the Clogston-Chandrasekhar limit (CCL), in which the magnetic energy associated with the spin susceptibility in the normal state exceeds the condensation energy in the superconducting state[19,20]. However, as the field is tilted away from the conducting planes, orbital mechanisms become feasible, producing an upper critical field that declines rapidly with increasing angle[11,14,16,17] The combination of the CCL and quasi-two-dimensional Fermi surface results in a very anisotropic critical field in the organics and copper oxides, strongly peaked for in-plane fields[11,14,16,17]. Even a cursory inspection of the data in Fig. 3 shows that this is definitely not the case in $(Ba,K)Fe_2As_2$.

The vortex structure is the second ingredient in the angle dependence of the resistive upper critical field in layered materials[9,11,14]. For superconductors with weakly-coupled layers, pancake vortices are observed for fields applied perpendicular to the quasi-two-dimensional planes, whereas Josephson vortices occur for in-plane fields. At intermediate orientations, the two types coexist, with their relative densities determined by the components of the field perpendicular and parallel to the layers. This results in a resistive transition with a shape that varies with field orientation, as seen in the organic superconductors and copper oxides[14,16,17]. To show that such behaviour does *not* occur in $(Ba,K)Fe_2As_2$, a single-crystal sample was rotated in-situ using a cryogenic goniometer, and the resistance measured at a fixed temperature of about 20 K for several field orientations. Data for the onset, midpoint and upper limit of the resistive transition are shown as a function of orientation in the inset b to Fig. 3. (Shown here are data for sample C, $Ba_{1-x}K_xFe_2As_2$ with $x \approx 0.4$ .) It is clear that the resistive transition varies in width and position very smoothly with field orientation, strongly suggesting that $(Ba,K)Fe_2As_2$ does not behave as a weakly-coupled layered superconductor, but instead that the vortex arrangement is similar at all field orientations.

The nearly isotropic critical field of $(Ba,K)Fe_2As_2$ might be linked to its distinctive Fermi surface. Published calculations[21], supported by recent magnetic quantum-oscillation measurements on its sister compound $SrFe_2As_2$[22], give a Fermi surface consisting of a number of strongly corrugated or flared tubes with relatively small (compared to the Brillouin zone) cross-sectional areas. Both calculated and measured effective masses are relatively light (~ a few times the free-electron mass) so that most likely orbital effects limit the upper critical field in $(Ba,K)Fe_2As_2$ for *H* parallel to *c*, as is seen in the copper oxides and organics. However, in contrast to the very weakly corrugated Fermi surfaces of the latter superconductors[15,18], that of $(Ba,K)Fe_2As_2$ shows considerable dispersion along the *c* direction in momentum space (denoted $k_c$), manifested as a strong oscillation of the cross-sectional area of the various sections as $k_c$ varies; the Fermi surface is much more *three dimensional*. The much larger corrugations in the Fermi surface of $(Ba,K)Fe_2As_2$ are sufficient to permit circulating currents *at all field orientations*; hence, orbital limiting effects might persist at all field angles, leading to the observed rather isotropic upper critical field.

Although the low-temperature upper critical field is rather isotropic, the initial slope of $H_{c2}(T)$ near $T_c$ does show some dependence on the field orientation (Fig. 3), perhaps resulting from details of the vortex structure (which will affect the magnetoresistance close to the transition[14]) or the Fermi surface topology. In our resistive critical field data, $dH_{c2}/dT(T=T_c)$ is determined to be about 5.4 T/K for $H//c$ and 2.9 T/K for $H \perp c$. These are close to the values found for $(Ba,K)Fe_2As_2$ in dc field measurements[7]. In contrast to the high $T_c$ copper oxides and the dirty $MgB_2$ in which a strong upturn curvature was observed in $H_{c2}(T)$ at very low temperatures[23,24], $H_{c2}(T)$ in $(Ba,K)Fe_2As_2$ shows a convex shape for $H$ perpendicular to c, but follows an almost linear temperature dependence down to 10 K for $H//c$. The strong, experimentally-measured, curvature of $H_{c2}(T)$ for $H$ perpendicular to c leads to a significantly lower zero temperature upper critical field compared to typical extrapolation methods; the latter yield values beyond 100 T, much higher than our experimentally inferred value of 70 T.

As far as we are aware, no other layered superconductors exhibit upper critical fields that behave in the same way as those of $(Ba,K)Fe_2As_2$. The difference is, we believe, associated with this material's distinctive Fermi-surface topology, the strong corrugations of which – a manifestation of essentially three-dimensional band structure - permit orbital limiting of the upper critical field at all field orientations. Therefore the 122-type ternary iron arsenides are unique in possessing both a rather high critical temperature and essentially three-dimensional electronic properties. In contrast to common assumptions based on the properties of the copper oxides, it seems that reduced dimensionality is not necessarily a prerequisite for "high-temperature" superconductivity.

*Note added in proof*: We have recently become aware that MHz penetration depth measurements in fields of up to 45T[27] further support our conclusion.

## References

1. Kamihara, Y., Watanabe, T., Hirano, M. & Hosono, H. Iron-based layered superconductor La[$O_{1-x}F_x$]FeAs (x = 0.05–0.12) with Tc = 26 K. *J. Am. Chem. Soc.* **130**, 3296–3297 (2008).

2. Chen, X. H. e*t. al*. Superconductivity at 43 K in SmFeAsO$_{1-x}$F$_x$. *Nature* **453**, 761-762 (2008).

3. Chen, G. F. e*t. al*. Superconductivity at 41 K and its competition with spin-density-wave instability in layered CeO$_{1-x}$F$_x$FeAs. *Phys. Rev. Lett*. **100**, 247002 (2008).

4. Ren, Z. A. *et. al*. Superconductivity in the iron-based F-doped layered quaternary compound Nd[$O_{1-x}F_x$]FeAs. *Europhys. Lett.* **82**, 57002 (2008).

5. Wang, C. *et. al.* Thorium-doping induced superconductivity up to 56 K in Gd$_{1-x}$Th$_x$FeAsO. *Europhys. Lett*. **83**, 67006 (2008).

6. Rotter, M., Tegel, M., and Johrendt, D. Superconductivity at 38 K in the iron arsenide (Ba$_{1-x}$K$_x$)Fe$_2$As$_2$. *Phys. Rev. Lett*. **101**, 107006 (2008).

7. Ni, N. *et. al*. Anisotropic thermodynamic and transport properties of single crystalline (Ba$_{1-x}$K$_x$)Fe$_2$As$_2$ (x = 0 and 0.45). *Phys. Rev. B* **78**, 014507 (2008).

8. Anderson, P. W. The theory of superconductivity in the high-Tc cuprate superconductors (Princeton University Press, Princeton NJ 1997).

9. See, e.g., Shrieffer, J. R. & Brooks, J. S. *Handbook of High-Temperature Superconductivity* (Springer, Berlin 2006).

10. Ding, H. *et. al.* Observation of Fermi-surface-dependent nodeless superconducting gaps in Ba$_{0.6}$K$_{0.4}$Fe$_2$As$_2$. *Europhys. Lett*. **83**, 47001 (2008).

11. Singleton, J. & Mielke, C. Quasi-two-dimensional organic superconductors: a review. *Contemp. Phys.* **43**, 63-96 (2002).

12. Huang, Q. *et. al.* Magnetic order in $BaFe_2As_2$, the parent compound of the FeAs based superconductors in a new structural family. Preprint at <http://arxiv.org/abs/0806.2776> (2008).

13. Riggs, S. C. *et. al.* Log-T divergence and insulator-to-metal crossover in the normal state resistivity of fluorine doped $SmFeAsO_{1-x}F_x$. Preprint at <http://arxiv.org/abs/0806.4011> (2008).

14. Nam, M. S. *et. al.* Angle dependence of the upper critical field in the layered organic superconductor kappa -$(BEDT-TTF)_2Cu(NCS)_2$(BEDT-TTF identical to bis(ethylene-dithio)tetrathiafulvalene). *J. Phys.: Condens. Matter* **11**, L477-484 (1999).

15. Singleton, J. *et. al.* Persistence to high temperatures of interlayer coherence in an organic superconductor. *Phys. Rev. Lett.* **99**, 027004 (2007).

16. Vedeneev, S. I. *et. al.*Reaching the Pauli limit in the cuprate $Bi_2Sr_2CuO_{6+\delta}$ in high parallel magnetic field. *Phys. Rev. B* **73**, 014528 (2006).

17. Li, P. C., Balakirev, F. F. & Greene, R. L. Upper critical field of electron-doped $Pr_{2-x}Ce_xCuO_{4-\delta}$ in parallel magnetic fields. *Phys. Rev.* **B 75**, 172508 (2007).

18. Sebastian, S. E. *et. al.* A multi-component Fermi surface in the vortex state of an underdoped high $T_c$ superconductor. *Nature* **454**, 200-203(2008).

19. Clogston, A. M. Upper limit for the critical field in hard superconductors. *Phys. Rev. Lett.* **9**, 266 (1962).

20. Chandrasekhar, B. S. A note on the maximum critical field of high-field superconductors. *Appl. Phys. Lett.* **1**, 7 (1962).

21. Liu, C. *et. al*. The Fermi surface of $Ba_{1-x}K_xFe_2As_2$ and its evolution with doping. Preprint at <http://arxiv.org/abs/0806.3453> (2008).

22. Sebastian, S. E. *et. al*. Quantum oscillations in the undoped parent magnetic phase of a high temperature superconductor. *J. Phys.: Condens. Matter* **20**, 422203 (2008) .

23. Gurevich, A. Limits of the upper critical field in dirty two-gap superconductors. *Physica C* **456,** 160–169 (2007).

24. Ando, Y. *et. al*. Resisitive upper critical fields and irreversibility lines of optimmally doped high $T_c$ cuprates. *Phys. Rev. B* **60**, 12475-12479 (1999).

25. Hunte, F. *et al*. Two-band superconductivity in $LaFeAsO_{0.89}F_{0.11}$ at very high magnetic fields. *Nature* **453**, 903-905 (2008).

26. Chen, G. F. *et al*. Breaking rotation symmetry in single crystal $SrFe_2As_2$. Preprint at <http://arXiv.org/abs/0806.2648> (2008).

27. Altarawneh, M. *et al*. Determination of anisotropic $H_{c2}$ up to 45 T in $(Ba_{0.55}K_{0.45})Fe_2As_2$ single crystals. Preprint at <http://arxiv.org/abs/0807.4488> (2008).

**Acknowledgements:** We acknowledge S. Riggs and J. Betts for experimental assistance and F. C. Zhang, Paul Goddard and Steve Blundell for fruitful discussions. Work at NHMFL-LANL is performed under the auspices of the National Science Foundation, Department of Energy and State of Florida. The experiments reported here are supported by the DOE BES program "Science in 100T", the NHMFL-UCGP, the National Science Foundation of China, the National Basic Research Program of China (973 Program) and the Chinese Academy of Sciences. H.Q.Y is also supported by PCSIRT of the Ministry of Education of China.

**Author Contributions** H. Q. Y. initiated and organized this project, did most of the experiments and analyzed the data. F. F. B. and J. S. provided experimental supports. S. A. B. measured the angle dependence of $H_{c2}$. Samples were provided by G. F. C., J. L. L. and N. L. W. The manuscript was drafted by H. Q. Y. and J. S.

# FIGURE CAPTIONS

**Figure 1: In-plane electrical resistivity of single crystal (Ba,K)Fe$_2$As$_2$ measured in pulsed high magnetic fields. a:** The temperature dependence of the resistivity $\rho_{ab}(T)$ at zero magnetic field. Note that there is a weak kink around 108 K in $\rho_{ab}(T)$ which may correspond to a spin-density-wave (SDW) or structural transition[7,12]. This indicates that the sample is located in the underdoped side. The inset shows the temperature dependence of the Hall coefficient $R_H$ obtained in pulsed magnetic fields of up to 60T. **b** and **c:** The field dependence of the resistivity $\rho(B)$ at various temperatures for fields parallel and perpendicular to the c-axis, respectively. In contrast to the broad superconducting-to-normal transitions observed in the 1111-type polycrystals[13,25] and some "high-$T_c$" cuprates[9], the single crystal (Ba,K)Fe$_2$As$_2$ grown by high temperature solution method[26] shows sharp transitions even in high fields, permitting accurate evaluations of the resistive upper critical field $H_{c2}$. In order to minimize the self heating effect in a pulsed magnetic field, samples with typical size of about 2mm×0.5mm×0.02mm were cleaved off from the big batch for resistivity measurement and no obvious heating effects were observed judging from the almost identical resistance curves collected in the up-sweeping and down-sweeping of the magnet. Longitudinal resistivity and transverse Hall resistivity were simultaneously measured with a typical 5-probe method in pulsed fields of up to 60 T at Los Alamos National High Magnetic Field Laboratory. Forward and reverse-field shots were made at the same temperatures for Hall resistance measurements in order to eliminate the effects of contact asymmetries. The data traces were recorded on a digitizer using a custom high-resolution low-noise synchronous lock-in technique. The temperature dependence of the resistivity was measured with a Lakeshore resistance bridge.

**Figure 2: Electrical resistivity versus temperature at selected magnetic fields. a:** $H//c$; **b:** $H\perp c$. Different symbols represent the resistivity $\rho_{ab}(T)$ at different magnetic fields, as labelled in the insets. The solid lines are guides to the eye and the dashed line in Fig. 2**a** highlights the upturn in the resistivity at low temperatures for $H//c$. The insets are the expansions of the superconducting sections. Note that besides the main superconducting transition there is an additional lower temperature/lower field transition in Figures 1b and 2b, suggesting the presence of a small amount of a second structural or chemical phase with a lower $T_c$. The amount (or visibility) of the second phase develops over time and with repeated thermal cycling of the samples; it is absent from "virgin" crystals. Note that Figures 1b and 2b were recorded several days after Figures 1a and 2a, with a number of thermal cycles between room- and cryogenic temperatures. The visibility of the lower-temperature transition increased with each thermal cycling.

**Figure 3: The upper critical field $H_{c2}(T)$.** The main figure shows $H_{c2}$ versus temperature for magnetic fields parallel to the c-axis (circles) and perpendicular to the c-axis (squares), in which the critical fields $H_{c2}(T)$ are determined from the midpoint of the sharp resistive superconducting transitions. Remarkably, the two samples (#A and #B) behave nearly identically. Inset **a:** The anisotropy parameter $\gamma = H_{c2}^{H \perp c} / H_{c2}^{H // c}$ plotted as a function of temperature. The parameter $\gamma$ is about 2 near $T_c$, a value close to that derived by other groups using low-field measurements[7], but decreases with decreasing temperature and approaches to 1 as $T \rightarrow 0$, indicating isotropic superconductivity. Inset **b:** The upper critical field $\mu_0 H_{c2}$ (sample #C, $T_c \approx 28$K), derived at the upper, midpoint and lower limits of the main resistive transition, is plotted as a function of the tilt angle $\theta$ at T=20 K, where $\theta$ is the angle between the applied magnetic field (**H**) and the crystallographic $c$ axis. The error bars mark the maximum deviation of the rounded transitions from the sharp part of the superconducting transition to the monotonically-increasing normal-state resistivity (upper limit), to the zero resistance (lower limit) and the width of the sharp transition (midpoint), respectively. For the angular dependence measurements, the electrical current is applied perpendicular to the direction of magnetic field. The angular dependence of $\mu_0 H_{c2}$ can be well scaled by $\varepsilon(\theta)H_{c2}=H_{c2}[\cos^2(\theta)+\gamma^{-2}\sin^2(\theta)]^{-0.5}$; the obtained effective mass anisotropy $\gamma$ (1.5$\pm$0.1) is very close to the value (1.4$\pm$0.1) we derived from $\gamma = H_{c2}^{H \perp c}/H_{c2}^{H // c}$ as shown in the inset **a**.

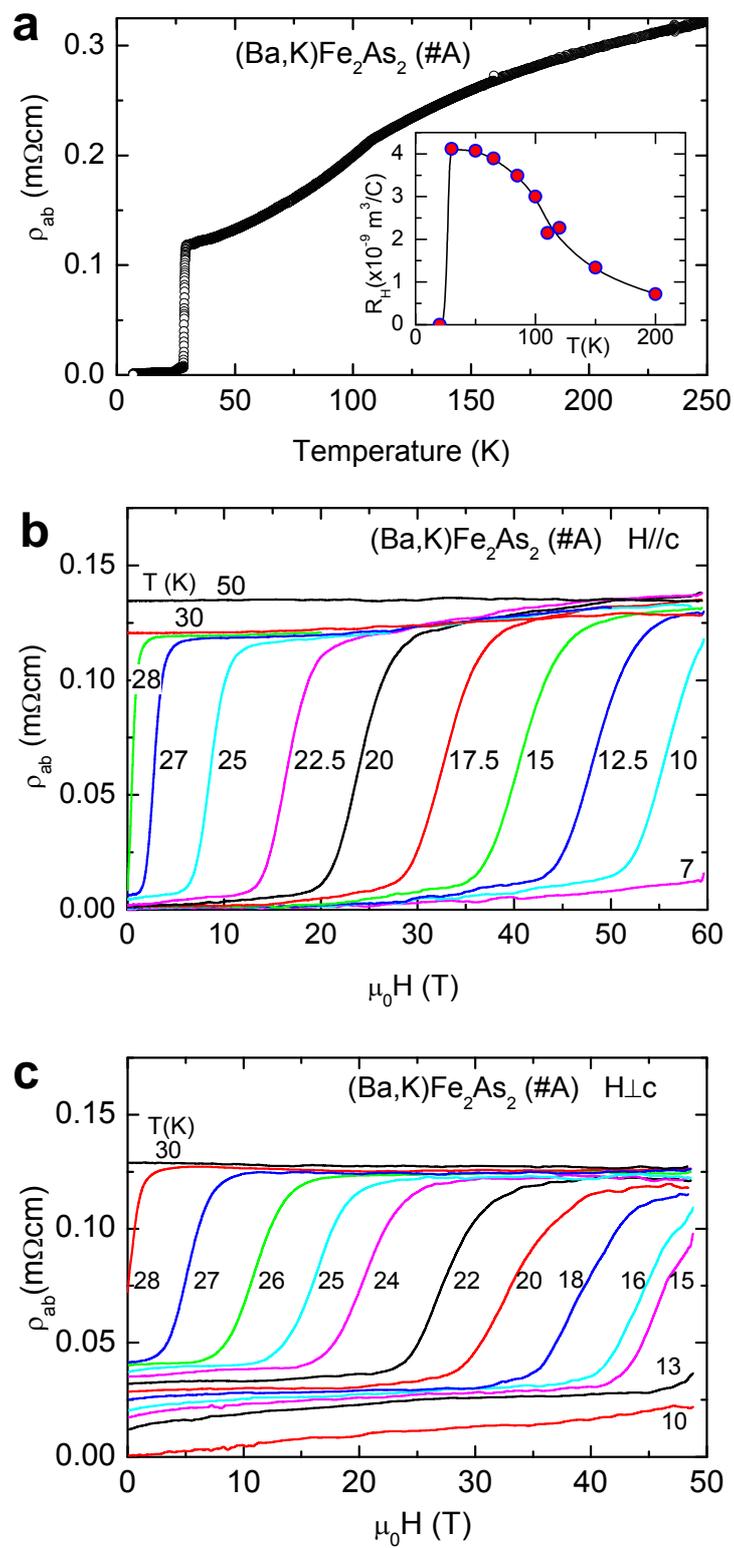

**FIGURE 1**

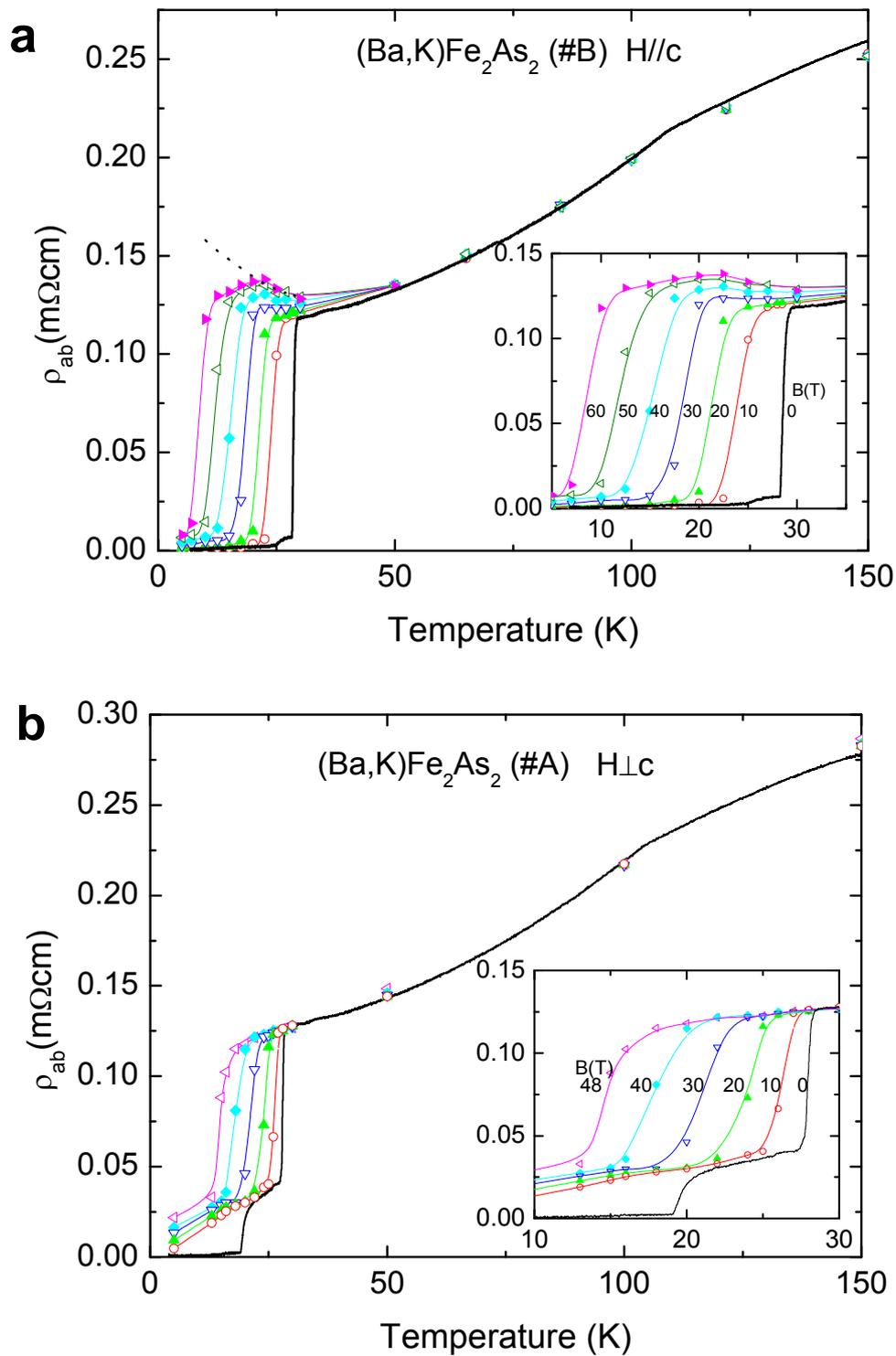

**FIGURE 2**

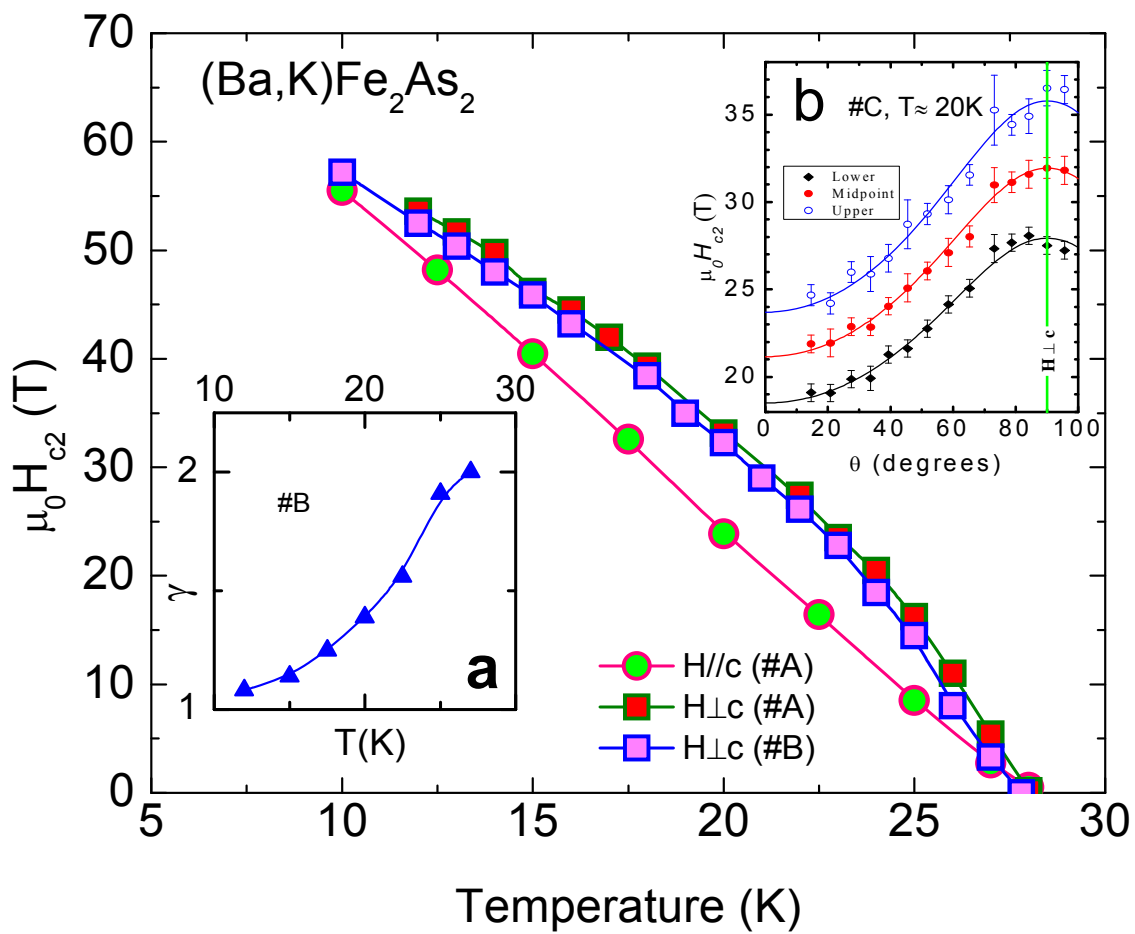

**FIGURE 3**